\documentclass{article}

\usepackage{PRIMEarxiv}

\usepackage[utf8]{inputenc} 
\usepackage[T1]{fontenc}    
\usepackage{hyperref}       
\usepackage{url}            
\usepackage{booktabs}       
\usepackage{amsfonts}       
\usepackage{nicefrac}       
\usepackage{microtype}      
\usepackage{lipsum}
\usepackage{fancyhdr}       
\usepackage{graphicx}       
\graphicspath{{media/}}     

\usepackage{listings}
\usepackage{xspace}
\usepackage[nopostdot,nogroupskip,nonumberlist]{glossaries}

\title{Scalable Quantum Optimisation using HADOF: Hamiltonian Auto-Decomposition Optimisation Framework
\thanks{\textit{\underline{Citation}}: Sankar, N., Miliotis, G. and Caton, S. Scalable Quantum Optimisation using HADOF: Hamiltonian Auto-Decomposition Optimisation Framework. In 
3rd International Workshop on AI for Quantum and Quantum for AI (AIQxQIA 2025), at the 28th European Conference on Artificial Intelligence (ECAI),
  October 25-30, 2025, Bologna, Italy}}

\author{
  Namasi G. Sankar \\
  School of Computer Science\\
  Centre for Quantum Engineering,\\ Science, and Technology\\
  University College Dublin \\
  Dublin\\
  \texttt{namasivayam.gomathisankar@ucdconnect.ie} \\
   \And
  Georgios Miliotis\\
    Antimicrobial Resistance and\\ Microbial Ecology Group, \\ School of Medicine, \\ University of Galway,  \\Galway\\
  \texttt{georgios.miliotis@universityofgalway.ie} \\
  \And
  Simon Caton \\
  School of Computer Science\\
  Centre for Quantum Engineering,\\ Science, and Technology\\
  University College Dublin \\
  Dublin\\
  \texttt{simon.caton@ucd.ie} \\
}

\begin{document}
\maketitle

\newacronym{approach}{HADOF}{Hamiltonian Auto-Decomposition Optimisation Framework}
\newcommand{\approach}{\gls{approach}\xspace}
\newacronym{qubo}{QUBO}{Quadratic Unconstrained Binary Optimisation}
\newcommand{\qubo}{\gls{qubo}\xspace}
\newacronym{qaoa}{QAOA}{Quantum Approximate Optimisation Algorithm}
\newcommand{\qaoa}{\gls{qaoa}\xspace}
\newacronym{tsp}{TSP}{Travelling Salesman Problem}
\newcommand{\tsp}{\gls{tsp}\xspace}

\begin{abstract}
Quantum Annealing (QA) and \qaoa are promising quantum optimisation algorithms used for finding approximate solutions to combinatorial problems on near-term NISQ systems. Many NP-hard problems can be reformulated as Quadratic Unconstrained Binary Optimization (QUBO), which maps naturally onto quantum Hamiltonians. However, the limited qubit counts of current NISQ devices restrict practical deployment of such algorithms. In this study, we present the \approach, which leverages an iterative strategy to automatically divide the \qubo Hamiltonian into sub-Hamiltonians which can be optimised separately using Hamiltonian based optimisers such as \qaoa, QA or Simulated Annealing (SA) and aggregated into a global solution. We compare \approach-with Simulated Annealing (SA) and the CPLEX exact solver, showing scalability to problem sizes far exceeding available qubits while maintaining competitive accuracy and runtime.. Furthermore, we realize \approach for a toy problem on an IBM quantum computer, showing promise for practical applications of quantum optimisation. 

\end{abstract}


\newif\iffinal

\iffinal
  \newcommand\namasi[1]{}
  \newcommand\simon[1]{}
  \newcommand\georgios[1]{}
\else
  \newcommand\namasi[1]{{\color{red}[***Namasi: #1]}}
  \newcommand\simon[1]{{\color{blue}[***Simon: #1]}}
  \newcommand\georgios[1]{{\color{magenta}[***Georgios: #1]}}
\fi

\glsresetall
\section{Introduction}
\label{sec:intro}


The \qubo model provides a unified framework for formulating many combinatorial optimization problems—such as the \tsp, graph partitioning, and scheduling—which are often NP-hard and difficult to scale using classical exact solvers \cite{NP-hard-qubo}. While specialized heuristics exist (e.g., Lin–Kernighan for \tsp) \cite{Lin-Kernighan}, they lack generality. In contrast, general-purpose \qubo solvers, including IBM Cplex \cite{docplex2024} and other MILP/QP engines \footnote{GNU Linear Programming Kit, available at: \url{http://www.gnu.org/software/glpk/glpk.html}, last visited: \today }, offer flexibility but struggle with large instances \cite{NP-hard-qubo}. \qubo also has the advantage of being represented as a Hamiltonian naturally, which can then be optimised via quantum computing \cite{ising-qubo} and classical Simulated Annealing (SA) \cite{simulated_annealing}. 

Quantum algorithms theoretically provide a scaling advantage for certain optimisation problems over classical methods \cite{Abbas2024-ri, QA-advantage}. Some quantum \qubo algorithms include \qaoa \cite{qaoa} , Quantum Annealing (QA) \cite{Suzuki2009-fv} and Grover Adaptive Search (GAS) \cite{Gilliam2021-xw}. \qaoa and QA provide multiple approximately optimal solutions in parallel, by taking advantage of quantum superposition. This is useful for many applications as it allows the domain expert to choose the best fitting solution for their particular problem and also compare different solutions. 

However, current quantum devices in the NISQ era have a limited number of qubits and cannot (yet) be used for practical and scalable applications \cite{Gilliam2021-xw}. In this study, we propose the \approach, a framework for the automatic decomposition of a global Hamiltonian into sub-Hamiltonians, using an iterative optimisation process. The \approach framework can be used to scale up many \qubo based algorithms such as \qaoa, QA, Feedback-Based Quantum Optimisation (FALQON) \cite{falqon} and SA \cite{Suzuki2009-fv}. Algorithms which produce a probability distribution over the solution space, from which solutions can be sampled - where good solutions are more likely to be sampled (approximately) are compatible with \approach. 

\approach recovers more information from the sampling distribution, beyond merely the single best solution, enabling \approach to scale to problem sizes much larger than the available number of qubits. We demonstrate, through classical simulation of \qaoa within our framework, that \approach surpasses Cplex on \qubo instances out of reach under the same classical hardware conditions, producing multiple high-quality solutions concurrently. Moreover, we argue that on actual quantum hardware, \approach would exhibit even greater performance acceleration, combining quantum advantage with heuristic flexibility. Our results show promise for \approach both as a quantum‑inspired classical algorithm and as a scalable method on NISQ-era and future quantum devices.




\section{Related Work}
\label{sec:rw}

\qaoa~\cite{qaoa} and QA~\cite{Suzuki2009-fv} are foundational quantum methods for tackling \qubo problems. \qaoa is a variational, gate-based algorithm alternating between cost and mixer Hamiltonians, generating a probability distribution over solutions favoring low-cost solutions~\cite{ising-qubo, kim2025distributed}. QA, on the other hand, is an analog adiabatic method evolving a quantum system from an initial to the problem Hamiltonian. Both produce biased sample distributions over candidate solutions, offering practitioners flexibility when selecting among high-quality alternatives. However, NISQ hardware limits QAOA/QA to small problem sizes due to qubit and connectivity constraints~\cite{kim2025distributed}. This motivates hybrid and decomposition approaches.

In classical optimization, divide-and-conquer and decomposition heuristics are standard for scaling to large problems. General purpose solvers (e.g., IBM Cplex~\cite{docplex2024}) can struggle \qubo problems beyond hundreds of variables~\cite{NP-hard-qubo}, motivating decomposition and hybrid quantum-classical methods. The multilevel \qaoa of Maciejewski \textit{et al.}~\cite{maciejewski2023multilevel} splits a large QUBO into manageable sub-QUBOs that are solved iteratively or in parallel and then recombined. These techniques enable practical scaling and lay the foundation for distributed quantum optimization.

Recent strategies distribute or decompose QAOA across subproblems. Recursive QAOA (RQAOA)~\cite{bravyi2020obstacles} uses QAOA to iteratively fix qubits and shrink the problem, focusing quantum resources on the hardest sub-instances. The QAOA-in-QAOA (QAOA$^2$) and related parallel QAOA heuristics~\cite{qaoa2} decompose a large graph (e.g., MaxCut) into subgraphs, solve each with QAOA in parallel, and merge the results, exploiting high-performance computing (HPC) for scalability. Early approaches worked best on sparse or weakly coupled problems~\cite{kim2025distributed}, but dense QUBOs require advanced coordination to manage strong variable interactions.

The Distributed QAOA (DQAOA) framework~\cite{kim2025distributed} extends parallelization further. Large QUBOs are decomposed into sub-QUBOs, solved on quantum or classical resources in parallel, with an aggregation policy reconciling overlaps and correlated interactions. This iterative approach scales to large, dense QUBOs; for example, Kim et al. report \~99\% approximation ratios on 1,000-variable instances within minutes, outperforming prior methods in both quality and time-to-solutionon. DQAOA leverages quantum-centric HPC platforms to update a global solution iteratively, demonstrating that distributed computing augments quantum optimization for practical problem sizes.

\approach and DQAOA overcome standard QAOA scalability limits via decomposition. DQAOA relies on explicit partitioning and parallel aggregation, excelling on HPC or distributed platforms. \approach uses adaptive, iterative refinement with a probabilistic global view, reducing quantum requirements per step. While DQAOA is optimal for raw parallelism and wall-clock minimization, \approach provides efficient sequential scaling and solution diversity. Both frameworks represent the cutting edge of distributed quantum optimization, and hybrid approaches combining their strengths are promising future directions.

\section{Approach}

\subsection*{General Overview}

\approach proceeds iteratively::
\begin{enumerate}
    \item Encode the full \qubo as a Hamiltonian.
    \item Apply an optimisation algorithm (like \qaoa or QA) that produces a probability distribution to sample approximate solutions.
    \item Approximate sub-Hamiltonians using the marginal probability of the binary variables (qubits).
    \item Solve each sub-Hamiltonian iteratively.
    \item Aggregate sampled solutions from sub-Hamiltonians to guide the next iteration.
\end{enumerate}

\subsection*{Implementation Details}

\approach introduces a new problem decomposition method, leveraging probabilistic state information, using an iterative refinement mechanism similar to classical optimisation loops.

Let the global problem be represented in the standard \qubo form:
\begin{equation}
\min_{x \in \{0, 1\}^n} \quad x^T Q x
\end{equation}
where $Q \in \mathbb{R}^{n\times n}$ is an upper triangular cost matrix, and $n$ is the dimensionality of the binary decision variable $x$. The corresponding quantum Hamiltonian $H_Q$ encodes the QUBO in the computational basis. 

To scale the optimisation process, $H_Q$ is decomposed into a set of sub-Hamiltonians, each defined over a subset of the full variable set. Let $S_i \subset \{x_1,...,x_n\}$ denote the variables of subproblem $i$, with $\|S_i\| = k<<n$. The sub-Hamiltonian $H_i$ is defined by: 
\begin{equation}
H_i = \mathbb{E}_{x_{\bar{S}_i}} = P ( H_Q \big| x_{S_i}, x_{\bar{S}_i})
\end{equation}
Here, $\bar{S}_i$ denotes the complement of $S_i$, and $P(x_{\bar{S}_i})$ is a distribution over unsampled variables.  We use \( \mathbb{E}[x_k] = P(x_k) \) as the marginal probability that variable \( x_k \) is 1, estimated from previous iterations or prior knowledge. Ideally, this expectation is estimated using a weighted average over all states the rest of the \qubo can assume. In this study, \( \mathbb{E}[x_k] \) is approximated as the expected value of each qubit, by sampling it. 

\begin{figure}
    \centering
    \includegraphics[width=1\linewidth]{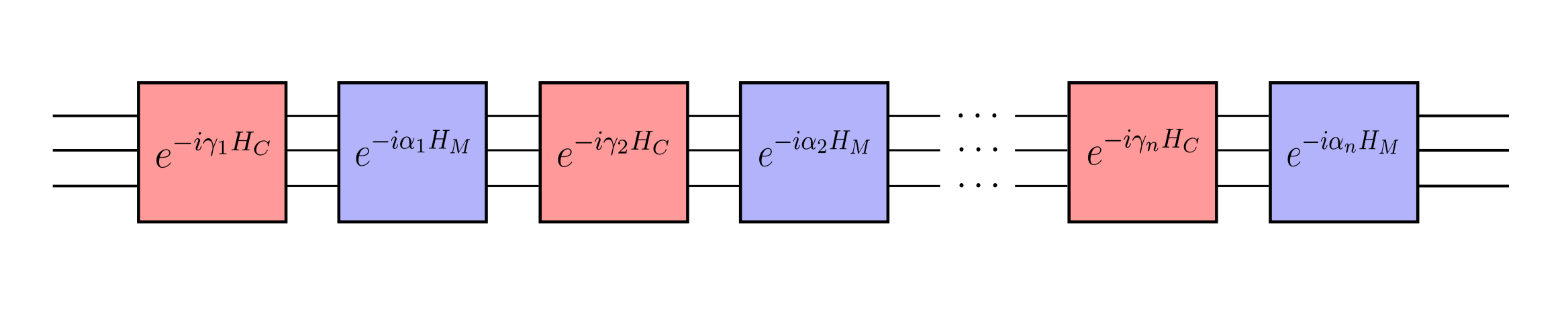}
    \caption{Standard \qaoa circuit with alternating cost and mixer Hamiltonians\cite{QUBO-pennylane}. The output produces a probability distribution over the solution space which can be samples with the shots parameter of the quantum simulation or backend. Higher sampled solutions are more likely to be solutions with better objective value.}
    \label{fig:QAOAcircuit}
\end{figure}

This transformation embeds global context into each subproblem while keeping the computational cost tractable. To estimate $\mathbb{E}[x_i]$, we use a modified \qaoa and SA procedure. We use the same $\beta$-schedule for both QAOA and SA.. For each subproblem \( i \), a \qaoa circuit is constructed using the cost and mixing unitaries, as in Figure \ref{fig:QAOAcircuit}:
\begin{align}
U(H_i, \gamma) &= e^{-m \gamma_m H_i} \\
U(M, \beta) &= e^{-m \beta_m \sum_{j=1}^{k} X_j}
\end{align}

\begin{figure}
    \centering
    \includegraphics[width=.6\linewidth]{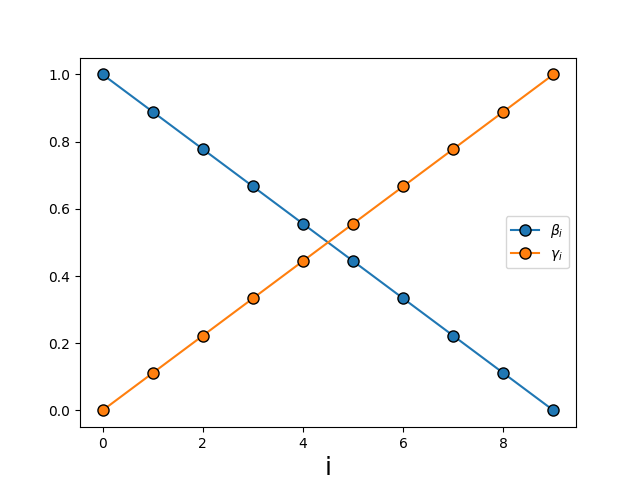}
    \caption{Trotterised \qaoa parameters based on Ref \cite{QUBO-pennylane}, \cite{openQAOA}.  We move $\beta_m$ from 1 to 0 and $\gamma_m$ from 0 to 1 allowing the system to stay close to the ground state of the mixer Hamiltonian to the Cost Hamiltonian.}
    \label{fig:qaoaparams}
\end{figure}

Here, we implement \qaoa as a trotterisation of QA, using the Annealing Parametrisation \cite{openQAOA}, to avoid the classical optimisation loop required to find $\beta_m$ and $\gamma_m$. We start in the ground state $\left| + \right\rangle^{\otimes k}$ of the mixer Hamiltonian $X$ and move to the ground state of the cost Hamiltonian $H_i$ slowly enough to always be close to the ground state of the Hamiltonian, as in Figure \ref{fig:qaoaparams}. This rate is determined by the number of layers $p$. We initialize the $\beta_m$ and $\gamma_m$ in this way, moving $\beta_m$ from 1 to 0 and $\gamma_m$ from 0 to 1. 

However, instead of applying the \qaoa procedure completely, two changes are made to iteratively estimate the sub-Hamiltonians. In each iteration of the loop, every sub-Hamiltonian is solved using \qaoa, however, the whole circuit is not applied. In the $l^{th}$ iteration, only layers 1 to $l$ are applied. 

\begin{figure}
    \centering
    \includegraphics[width=1\linewidth]{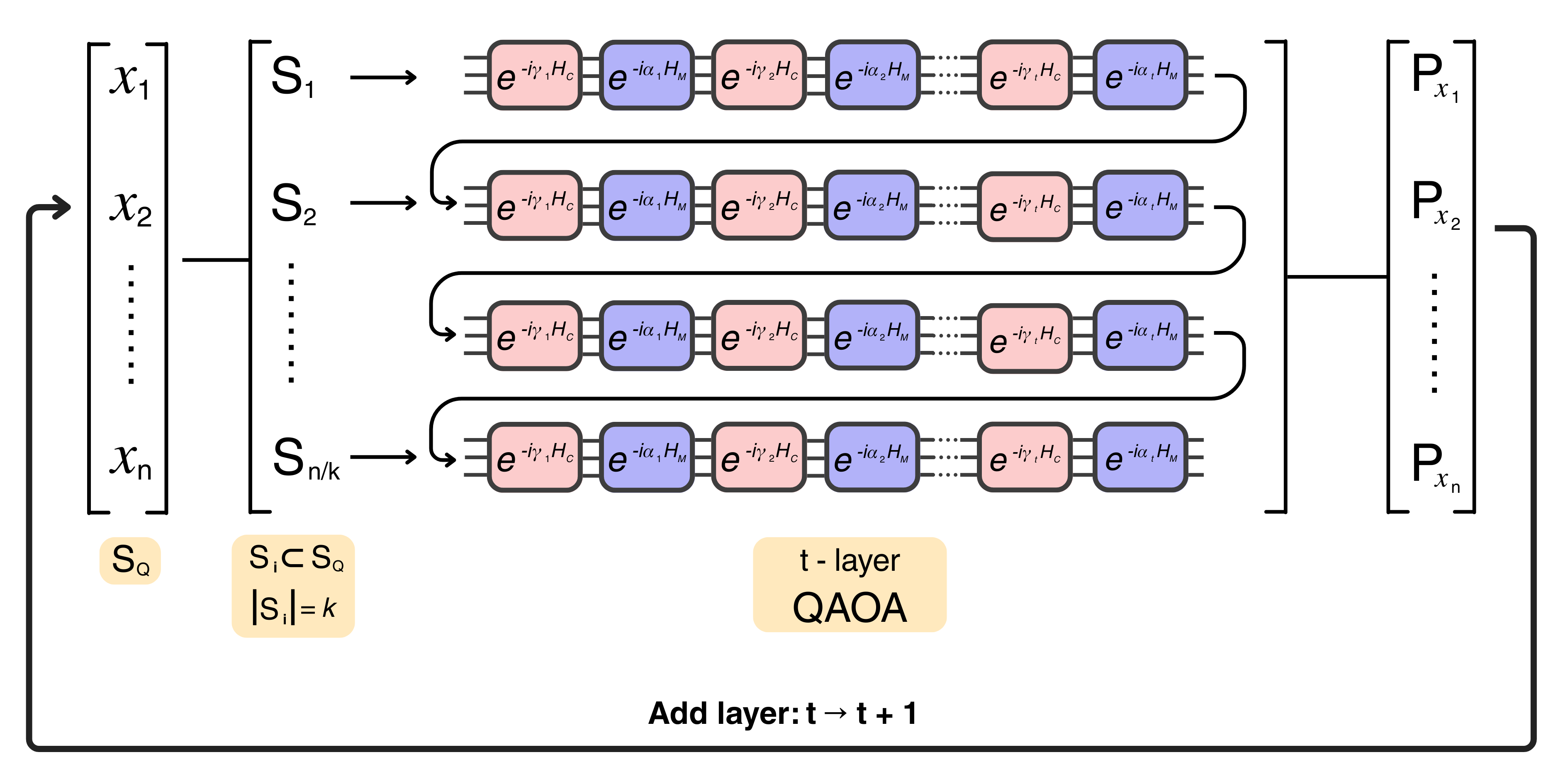}
    \caption{General overview of the \approach framework. Here, we use \qaoa as the optimiser, which is called iteratively. 
    \textbf{(1)} We choose subsets of sizes 5 and 10 from the binary variables of the global problem. 
    \textbf{(2)} These are used to form the sub-Hamiltonians using $P(x_i)$, approximated as the expected value of each qubit. 
    \textbf{(3)} The \qaoa circuit set up with $t=1$ layers and in every iteration we add a layer. In this study, we use 10 layers. 
    \textbf{(4)} Once the $n/k$ sub-Hamiltonians are optimised, we sample them and use an aggregation policy to form the global solution probability distribution.}
    \label{fig:HADOF}
\end{figure}

To approximate the value of $\mathbb{E}[x_i]$, individual qubits are sampled instead of sampling from all possible solutions of the \qaoa in every iteration. The average for each qubit is used as a proxy for $\mathbb{E}[x_i]$. We follow the same procedure for beta scheduling while using SA \approach. The optimisation process unfolds over \( p \) global iterations as in Figure \ref{fig:HADOF}. 

\section{Parametric Details}

\subsection*{Step-by-step Procedure (Pseudocode)}

\begin{enumerate}
    \item \textbf{Initialisation:} 
    \begin{itemize}
        \item Create a global probability vector \( P(x_k) \in \mathbb{R}^n \), with all values initialised to 0.5, except for the first subset $S_i$
    \end{itemize}
    
    \item \textbf{Iterative \qaoa Loop:}
    For \( l = 1, 2, \dots, p \):
    \begin{enumerate}
        \item Initialise \qaoa circuit with $l$ layers with the first $l$ values of $\beta_m$ and $\gamma_m$
        \item For each model \( i \in \{0, \dots, M-1\} \):
        \begin{itemize}
            \item Replace inactive variables $\bar{S}_i$ by fixed expected values from previous iterations
            \item Construct sub-QUBO for subset \( S_i \) using the expected values $P(x_{\bar{S}_i})$ from previous iteration.
            \item Convert to Ising form: \( Q \to (h, J) \) and get the sub-Hamiltonian corresponding to the sub-QUBO
            \item Apply \qaoa circuit of depth \( l \) on \( k \) qubits where \( k = \|S_i\|\) :
            \item Update \( P(x_k) \) vector by measure expected values of each qubit for current model:
            \begin{equation}
            P(x_i) = \mathbb{E}[x_i] 
            \end{equation}
        \end{itemize}
    \end{enumerate}
    
    \item \textbf{Final Output:} 
    After the final iteration, run full \qaoa with full depth - all the layers in the beta schedule - to extract binary samples. Collect and store final solutions from all models. These solutions and their probabilities can be aggregated to form the global solution.
\end{enumerate}

We generate \qubo problems by filling an upper triangular matrix using a uniform random number generator between -10 and 10. We present comparisons with CPLEX for problems with $n = 10, 20, ..., 100$ binary variables, and scale up to larger problems of size $n = 100, 200, ..., 500$ variables for the SA and \approach methods. We choose \( k = 5\) and \( k = 10\), where number of \qaoa  and SA circuits per iteration will be \( n/k \). 

We initialise $P(x_i) = 0.5$ for all $i$. Circuits use 10 layers with $\beta_m = 1-(m/10)$ and $\gamma_m = m/10$. After each sweep of the $n/k$ circuits, we add one layer. To measure the individual qubits to update $P(x_i)$ we use 500 shots per qubit. 

Finally, we sample each circuit over all $k$ qubits using 5000 shots per circuit, to produce a distribution over each sub-solution. In this study, we only aggregate the solutions in a rudimentary manner. We form 5,000 global solutions by concatenating sampled sub-solutions in sampling order and then evaluate their objective values.


\section{Results and Discussion}

\begin{figure}
    \centering
    \includegraphics[width=1\linewidth]{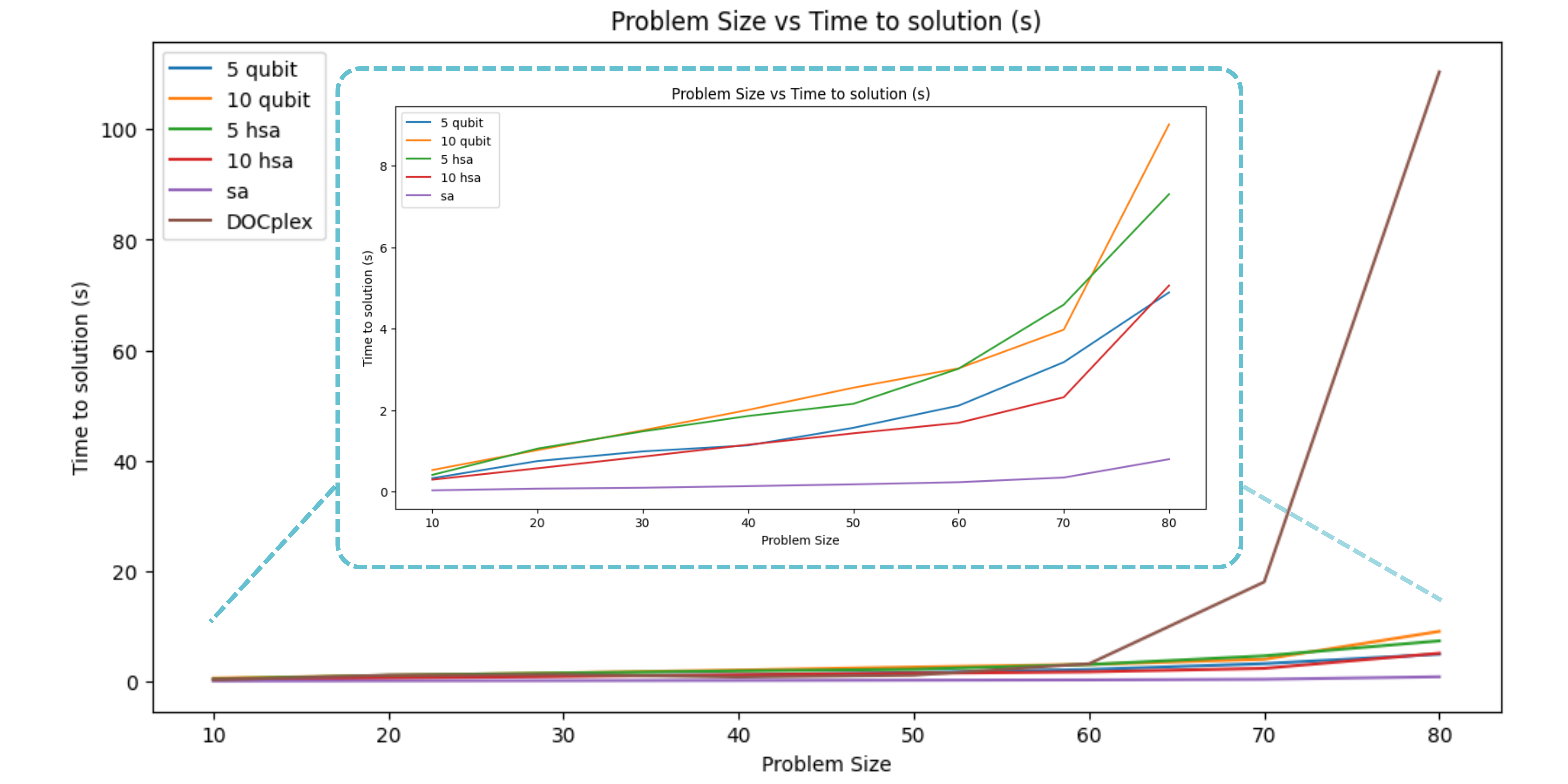}
    \caption{Time to solution as a function of problem size for CPLEX (exact classical solver), SA, HADOF QAOA and SA with 5 variable and 10 variable sub-QUBOs for problem sizes $n=10$ to $n=80$. CPLEX exhibits exponential scaling. SA scales the best in time as problem size increases. The inset shows the other algorithms, excluding CPLEX for clarity.}
    \label{fig:scaling-3}
\end{figure}

\begin{figure}
    \centering
    \includegraphics[width=0.92\linewidth]{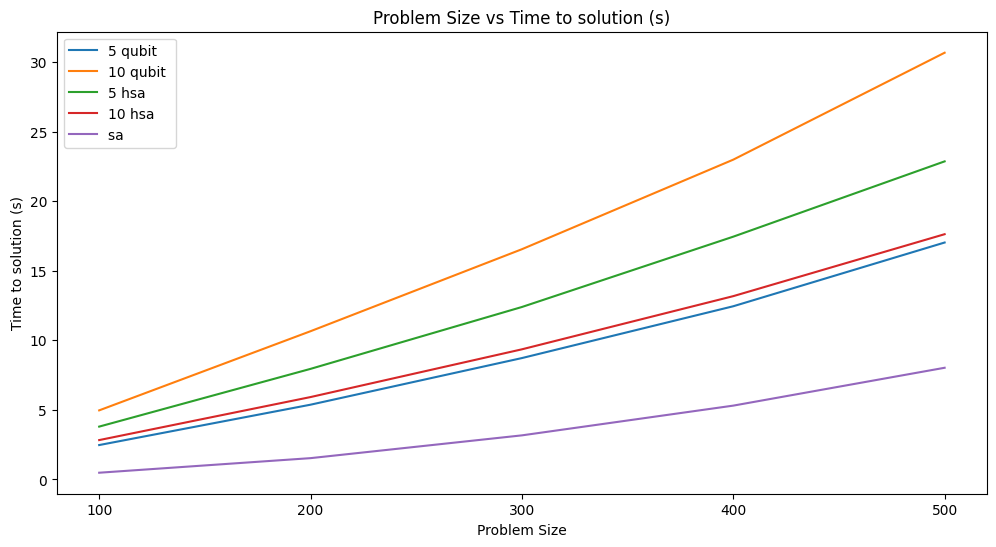}
    \caption{Time to solution as a function of problem size for SA, HADOF QAOA and SA with 5 variable and 10 variable sub-QUBOs for problem sizes $n=100$ to $n=500$.}
    \label{fig:500time}
\end{figure}

We evaluated \approach on randomly generated QUBO instances of varying sizes. We compared its performance with Pennylane \cite{pennylane} classically simulated \qaoa circuits, SimulatedAnnealingSampler from the D‑Wave Ocean SDK \footnote{D-wave ocean software, available at: \url{https://docs.ocean.dwavesys.com/}, last visited: \today} and the classical IBM Cplex solver \cite{docplex2024}.

For each problem size, we generated 100 independent QUBO instances. The 5 and 10 qubit \approach \qaoa, 5 and 10 variable \approach SA, global problem SA and global problem CPLEX methods were run on identical problem sets. This allows us to compare SA on the global problem directly against SA using \approach. The results were scaled such that the CPLEX objective value was set to one for problem sizes from 10-80. Beyond 80 variables, CPLEX became intractable and the solutions are scaled such that SA objective value is set to 1.  We perform all the simulations on an Apple M3 Pro device. We produce a distribution of 5000 sample solutions for each algorithm, except CPLEX. This allows us to calculate the average solution objective value of the distribution. We also define two individual solutions from these as best objective value and most probable objective value. The best objective value is the solution with the best objective from the 5000 global solutions. The most probable solution is defined as the aggregation of the most probable sub-solutions from each circuit. These values are used to compare the accuracies of the algorithms. 

\paragraph{Scalability and Runtime} 
Figure~\ref{fig:scaling-3} shows the \textit{time to solution} for CPLEX, SA and the 5-qubit and 10-qubit \approach approaches using \qaoa and SA for 10-100 variable problems. The classical CPLEX solver demonstrates exponential scaling with problem size, as expected for exact solvers on NP-hard combinatorial problems~\cite{NP-hard-qubo}. CPLEX is nearly instantaneous up to 40 variables, but runtime rapidly increases at larger sizes. In contrast, the four \approach approaches and SA display better scaling even upto problem sizes of 500 as shown in Figure~\ref{fig:500time}. This indicates that HADOF can outperform exact solvers like CPLEX in runtime for moderate sizes, even when \qaoa is classically simulated. The 5-qubit \qaoa version is consistently faster than the 10-qubit \qaoa. This could be because simulation of \qaoa classically is expensive as the circuit size increases. We see that the 10 variable (\approach SA) HSA is faster than the 5 variable \approach. We note that SA takes the least time to solve all of the problems.

\begin{figure}
    \centering
    \includegraphics[width=1\linewidth]{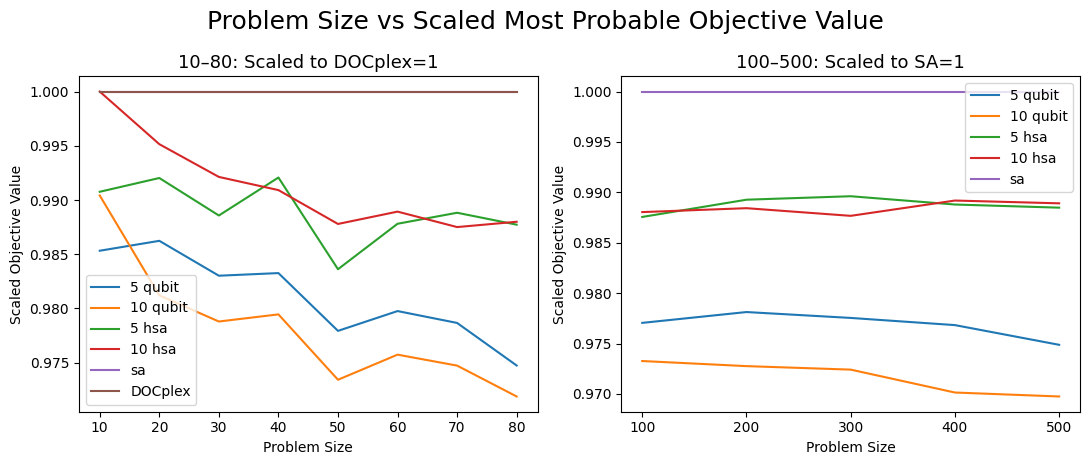}
    \caption{For small problems, we note that SA finds the exact solution offered by CPLEX all the time. The \approach based methods decrease in scaled objective value. However for larger problem sizes, they stay around 0.98 for HSA and around 0.975 for \qaoa, with the 5 qubit circuits performing better than the 10.}
    \label{fig:probobj}
\end{figure}

\begin{figure}
    \centering
    \includegraphics[width=1\linewidth]{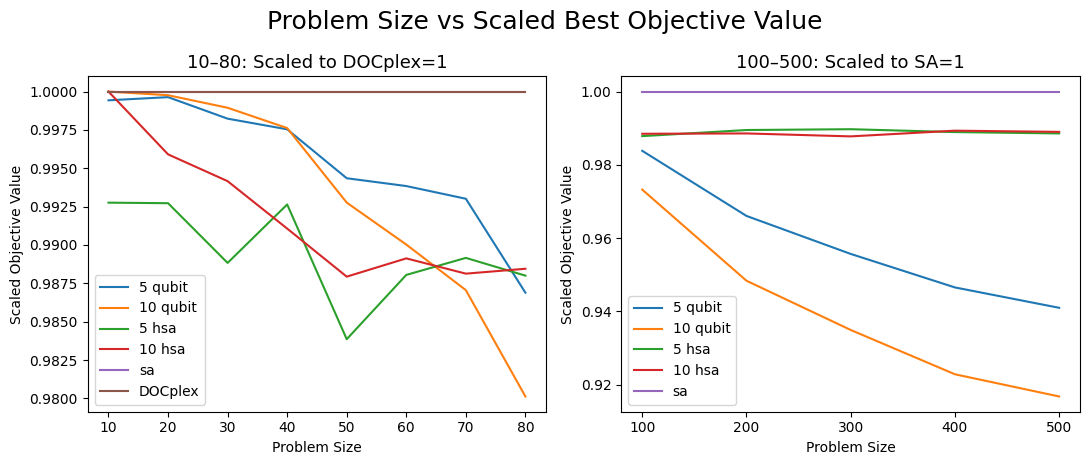}
    \caption{The \approach based methods decrease in scaled objective value for small problem sizes. However for larger problem sizes, they stay around 0.98 for HSA but keep reducing for \qaoa, with the 5 qubit circuits performing better than the 10 for all problem sizes.}
    \label{fig:bestobj}
\end{figure}

\begin{figure}
    \centering
    \includegraphics[width=1\linewidth]{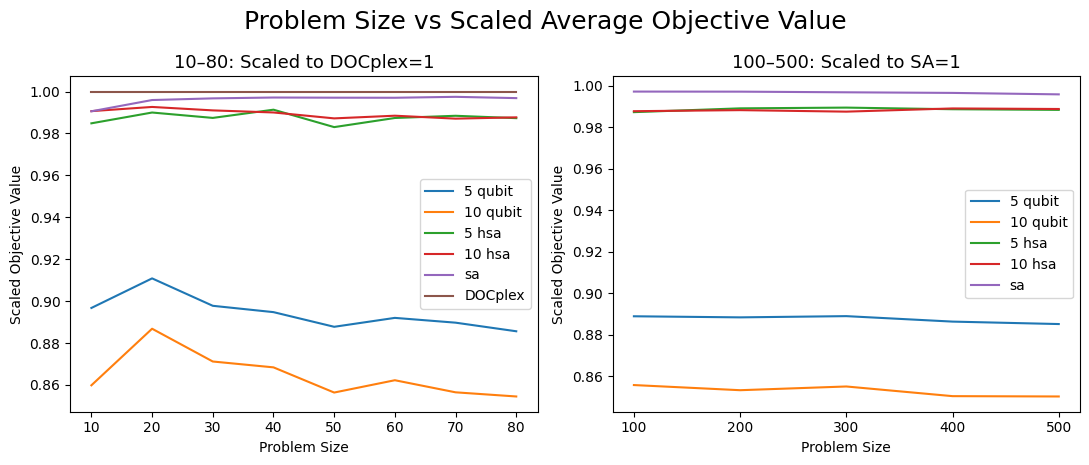}
    \caption{For small problems, we note that SA produces a distribution of 5000 solutions that is near optimal. The \approach based methods exhibit stable average objective values with above 0.85 for all methods.}
    \label{fig:avgobj}
\end{figure}

\paragraph{Solution Quality}
Figures~\ref{fig:probobj}, \ref{fig:bestobj}, and \ref{fig:avgobj} display the \textit{scaled objective values} for the most probable solutions, best solutions and average solutions across the algorithms respectively. These objective values are scaled to CPLEX solution for the 10-80 variable problems and scaled to SA solution for 100-500. Across all problems from 10-80, SA and CPLEX find the most optimal solution. The \approach methods initially decrease in accuracy of best and most probable solutions as the problem size increases (10-80), but their average accuracies tend to stay stable around 0.86 and 0.90 for the 10 and 5 qubit \qaoa. It stays above 0.98 using HSA. The sampling-based nature of HADOF preserves not only high solution quality but also solution diversity, as in SA, which is valuable in practical combinatorial settings~\cite{qaoa}.

\paragraph{Modularity}
\approach is a framework that uses an optimisation process within it, to scale up the problem sizes that can be solved by it. In this research, we tested it using SA and \qaoa. The framework requires that the algorithm produces a probability distribution over the solution space, from which solutions can be sampled - where good solutions are more likely to be sampled. Similar algorithms such as QA and FALQON \cite{falqon} may be compatble with \approach as well.

\paragraph{Testing on a Real Device}
We generated a single 20-variable QUBO and executed \approach \qaoa on IBM's cloud-accessible quantum device through QiskitRuntimeService \cite{qiskit-ibm-runtime} using 5-qubit circuits. Using the same beta scheduling parameters with 10 layers resulted in a solution with 0.84 objective value of the CPLEX solution. It took 6m and 42s to run including the classical calculation of sub-Hamiltonians and the queueing time on the real device. The Pennylane circuits were directly executable by changing the backend. Further rigourous evaluation is required to understand how \approach performs on real NISQ devices, and with larger problem sizes.

\paragraph{Summary}
\approach achieves hardware-efficient optimization by requiring only small quantum circuits regardless of global problem size, scaling to $n=500$ and beyond. The framework delivers not only near-optimal objective values but also a diversity of high-quality solutions, thanks to its iterative and sampling-based design. \approach is also modular and may be able to improve the scalability of many different algorithms.

\section{Conclusion and Future Work}


We introduced \approach, a Hamiltonian Auto-Decomposition Optimization Framework, and demonstrated its capability to solve large-scale \qubo problems efficiently by iteratively dividing them into tractable subproblems. Our results show that \approach outperforms the classical CPLEX solver in runtime and scalability, solving problems up to 500 variables that are otherwise infeasible for CPLEX under the same hardware constraints. Notably, \approach maintains near-optimal solution quality and delivers multiple high-quality solutions in a single run.

\approach offers several potential advances for quantum optimisation. It is extremely hardware efficient by taking advantage of only \( k << n \) qubits at any time to explore a high-dimensional \qubo space, allowing NISQ based algorithms to explore large problems irrespective of qubit availability. It yields a distribution of high-quality solutions instead rather than a single optimum. Another interesting perspective of exploration could use \approach to understand \qubo decomposition, as it iteratively creates sub-QUBOs that can be optimised separately and aggregated. This could be useful for improving or parallelising even classical algorithms which produce a distribution of solutions such as SA. 

\approach is also modular with many different optimisation algorithms that can be used under the framework to scale up beyond the available number of qubits or other device limitations that restrict the number of variables that can be solved at once.

Another key finding is that \approach based \qaoa is highly scalable even while simulating it on a classical device. We are able to solve large size problems beyond classical solver limits (e.g., Cplex) on the same machine, in simulation. Real device implementation of \approach may show speedups even over fast and approximate classical algorithms like SA \approach based SA. It would be useful to study how \approach fares against classical approximate and heuristic solvers.

While our simple aggregation—combining subproblem samples—was sufficient to surpass classical solvers in some regimes, future work will develop more robust policies to assemble sub-Hamiltonian samples into a coherent global distribution. We anticipate that adopting ideas from distributed QAOA (DQAOA)~\cite{kim2025distributed} and multi-level frameworks \cite{maciejewski2023multilevel}, such as adaptive coarse-to-fine decomposition and weighted aggregation, will allow us to better capture variable dependencies and further improve global sampling.

Following a similar sub-problem selection and aggregation strategy as in DQAOA \cite{kim2025distributed} may help parallelise \approach to run on multiple classical or quantum cores simultaneously. This ability to run large problems using small circuit sizes in embarrassingly parallel loops may allow us to further speed up and scale up the problems we can solve on current NISQ hardware.

Our results are based on simulated quantum circuits. Validation on real gate-based and annealing hardware is needed to quantify potential advantages in scalability and speed. It is important to quantify how the algorithm is affected under noisy NISQ hardware. 

In addition, \approach's decomposition scheme can be leveraged as a general divide-and-conquer technique for large QUBO problems. We plan to explore its use as a modular component within hybrid quantum-classical solvers, extending its scalability to industry-scale optimization. As quantum hardware advances, deploying \approach with larger sub-circuits will also be investigated. Ultimately, our goal is to integrate enhanced aggregation strategies and multi-level learning to realize a fully scalable quantum-classical hybrid solver capable of addressing practical, large-scale combinatorial optimization.

\bibliographystyle{unsrt}  
\bibliography{refs}

\end{document}